\documentstyle[preprint,aps,epsfig]{revtex} 

\newcommand{\be}{\begin{equation}}
\newcommand{\ee}{\end{equation}}
\newcommand{\bea}{\begin{eqnarray}}
\newcommand{\eea}{\end{eqnarray}}

\newcommand{\eett}{e^- e^+ \to t \bar{t} }

\begin{document}

\draft

\preprint{
\begin{tabular}{l}
\hbox to\hsize{November, 2000\hfill KIAS-P00071}\\[-2mm]
\hbox to\hsize{           \hfill SNUTP 00-031}\\[-2mm]
\hbox to\hsize{           \hfill hep-ph/0011173}\\[-3mm]
\end{tabular} 
}

\title{
Probing $Z^\prime$ gauge boson with
the spin configuration of top quark pair production
at future $e^- e^+$ linear colliders
} 

\author{
Kang Young Lee$^1$,
Seong Chan Park$^2$,
H. S. Song$^2$,
and Chaehyun Yu$^2$
}

\address{
$^{1}$ School of Physics, Korea Institute for Advanced Study, 
Seoul 130-012, Korea\\
$^{2}$ Department of Physics,
Seoul National University, Seoul 151-742, Korea \\
}

\date{\today}
\maketitle

\begin{abstract}

We explore the effects of extra neutral gauge boson 
involved in the supersymmetric $E_6$ model 
on the spin configuration of the top quark pair produced 
at the polarized $e^- e^+$ collider.
Generic mixing terms are considered 
including kinetic mixing terms as well as mass mixing.
In the off-diagonal spin basis of the standard model,
we show that the cross sections for the suppressed 
spin configurations can be enhanced 
with the effects of the $Z'$ boson 
through the modification of
the spin configuration of produced top quark pair 
enough to be measured in the Linear Colliders,
which provides the way to observe the effects
of $Z'$ boson and discriminate the pattern of
gauge group decomposition. 
It is pointed out that the kinetic mixing 
may dilute the effects of mass mixing terms,
and we have to perform the combined analysis.

\end{abstract}


\narrowtext

\section{Introduction}

The existence of extra U(1) gauge group is inevitable 
in many extended models derived by
grand unified theories (GUTs) of higher ranked gauge group
and superstring/M theories inspired models \cite{extraz}.
Moreover, the symmetry breaking scale for extra U(1) might be
as low as ${\cal O}(1)$ TeV in the context of supersymmetry 
to provide the solution of the $\mu$ problem \cite{muproblem}, 
leading to the possibility to observe the effects of 
the extra heavy neutral gauge boson, $Z'$ boson
at the future collider experiments. 
The latest bound of the $Z'$ boson mass comes from
a direct search at the $p \bar{p}$ collider via Drell--Yan production
and subsequent decay to charged leptons \cite{CDF}.
Indirect constraints for the $Z'$ boson mass and
the $Z-Z'$ mixing angles are obtained from 
high precision LEP data at the $Z$ peak energy and 
from various low energy neutral current experiment data
\cite{erler,chay,langacker,indirectz}.
Very recently it is indicated that 
a missing invisible width in $Z$ decays at LEP 1 and
a significantly negative $S$ parameter observed 
in atomic parity violation of Cs atom can be explained properly 
if the presence of an $Z^\prime$ boson is assumed \cite{apv}.
Thus it is timely and interesting to search for 
an effective way to probe the 
effects of $Z'$ boson at future colliders.  

As a larger class of $Z'$ models are considered 
from the string perspective, meanwhile,
it is natural to introduce kinetic mixing term.
The kinetic mixing is a threshold effects of string models
at the string scale and can be generated 
by the renormalization group (RG) evolution 
from the high energy scale to the scale that we study. 
Furthermore it may yield significant effects
on the phenomenologies of the $Z'$ couplings to the SM sector
\cite{babu}.

In this paper, we study the effects of $Z'$ boson
involved in the supersymmetric E$_6$ model
in $\eett$ process at linear colliders (LC) 
including the kinetic mixing term.
Charges of the standard model (SM) fermions for $Z'$ boson
are determined by the gauge group decompositions,
along with which the $\psi$-, $\chi$-, and $\eta$-models
are defined at low energy scale in the SUSY E$_6$ model framework.
Here we take the decoupling limit of the exotic fermions.
Since the asymmetry of the left- and right-handed couplings to
$Z'$ boson characterizes each model,
it is possible to distinguish models
using the spin informations of top quark pair
with the polarized initial $e^- e^+$ beams.
We can read out the information of the polarization of top quark 
through the angular distribution of the decay products 
\cite{spin-correlation}.
The large mass of the top quark prompts itself to decay before 
hadronization and the information of the top spin 
is free from the uncertainties of hadronization.
The top quark pair is produced in an unique spin configuration
at the polarized $e^- e^+$ collision,
which reveals remarkable features \cite{parke1,lee}
in the off-diagonal basis of spin. 
Moreover it is interesting to observe the polarization of top quark 
to probe new physics in this basis,
since the off-diagonal basis is model-dependently defined.

The LC with $\sqrt{s}=500$ GeV is
the best testing ground for studying the $t \bar{t}$ production 
in the off-diagonal basis.
If the CM energy is around at threshold of 
top pair production, the top spins are determined 
by the electron and positron momentum directions
since top quark pair is almost at rest.
Then the off-diagonal basis cannot be defined.
At high energy, $\sqrt{s} \gg m_t$,
the spin basis is close to the usual helicity basis
so that the angle $\xi \sim 0$.
Thus it is hard to extract the new physics effects
from $\xi$ although they exist.

This paper is organized as follows. 
In section II, the extra neutral gauge bosons 
in the string inspired supersymmetric E$_6$ are briefly reviewed. 
In section III, we present the formulae of scattering amplitudes 
of $\eett$ process with the $Z'$ bosons 
in a generic spin basis of the top quark pair.
The off-diagonal spin basis is defined and 
discussed is the way how to probe the $Z'$ boson 
through the spin configuration of $t \bar{t}$ pair.
In section IV, the numerical analysis for each models are performed
in the SM off-diagonal basis. 
Section V is devoted to summary of the paper.

\section{$Z^\prime$ boson in supersymmetric E$_6$ model}

In the supersymmetric E$_6$ model, 
there are two additional U(1) factors beyond the SM gauge group 
since the rank of E$_6$ group is 6.
The canonical decompositions of $\psi$ and $\chi$ models are as follows:
\bea
{\rm E}_6 \rightarrow &{\rm SO(10)}&\times {\rm U(1)}_{\psi}, 
\nonumber \\
               &{\rm SO(10)}& \rightarrow {\rm SU(5)} \times {\rm U(1)}_{\chi}.
\nonumber 
\eea  
After the extra U(1) symmetries are spontaneously broken 
by the weak iso-singlet Higgs scalar(s), 
the gauge bosons $Z_{\psi}$ and $Z_{\chi}$ corresponding to 
the groups U(1)$_{\psi}$ and U(1)$_{\chi}$ respectively 
become massive but are not mass eigenstates in general. 
We call it $Z'$ a linear combination of $Z_{\psi}$ and $Z_{\chi}$
parametrized by the mixing angle ${\theta}_E$ 
\be 
Z^\prime ({\theta}_E) \equiv 
    Z_{\chi} \cos {\theta}_E + Z_{\psi} \sin {\theta}_E, 
\ee
which is relatively light enough to mix with the ordinary $Z$ boson 
and relevant to the low energy phenomenology.
The orthogonal mode to $Z'$ boson is assumed to be so massive 
that its effect is to be decomposed. 
In the case of ${\theta}_E =0$, the $Z^\prime$ mode is 
identified to $Z_{\psi}$ boson; 
if ${\theta}_E =\pi /2$, $Z^\prime$ mode is $Z_{\chi}$ boson. 
The $\eta$-model and corresponding $Z_{\eta}$ boson
is defined by setting ${\theta}_E =\tan^{-1} (-\sqrt{5/3})$.
Here we assume that exotic fermions are heavy enough to
be decoupled.

In the effective rank-5 limit with only one extra neutral gauge boson,
the interaction Lagrangian is described by
\be 
-{\cal L}_{\rm int} = \sum_{f} \bar{\Psi}_f  {\gamma}^{\mu}
                      \bigg[  g_{3} {\lambda}^{\alpha} G^{\alpha} _\mu 
                       + g_{2} {T^a}_f {W^a} _{\mu}
                       + g_{1} Y_f B_{\mu}
                       + g'_{1} {1 \over 2} 
         ({f^f}_{\rm V} -{{f^f}_{\rm A}}{\gamma ^5} ){Z}^{\prime}_{\mu}\bigg]
                      {\Psi}_f, 
\ee
where 
$\Psi _f$ is the fermion field with flavour $f$; 
${\lambda}^{\alpha}$ and $T^a$ are generators 
of $SU(3)_{C}$ and $SU(2)_{L}$ gauge group respectively.
The extra gauge coupling is expressed by  
$g_1^\prime \equiv {1 \over \sqrt \lambda} g_1$ 
with order 1 parameter  $\lambda $.  
The exact value of $\lambda$ depends upon the pattern of symmetry breaking 
and we set the value 1 in the numerical analysis. 
The couplings, ${f^f}_{V}$ and ${f^f}_{A}$ are the vector and
axial vector charges of the fermion for U(1)$^\prime$ group. 
The U(1)$^\prime$ charge assignment is given in the Table 1
in terms of the vector and axial vector couplings of $Z'$ to fermions 

\begin{center}
\begin{tabular}{|c|c|c|c|c|c|c|}
\hline
Models
& \multicolumn{2}{c|}{$\psi$}
& \multicolumn{2}{c|}{$\chi$}
& \multicolumn{2}{c|}{$\eta$} \\ \cline{2-3}\cline{4-5}\cline{6-7}
{particles}& $f_{V}/\sqrt{5/72}$ & $f_{A}/\sqrt{5/72}$ 
& $ ~~2 \sqrt{6}f_{V}$~ & ~~$  2 \sqrt{6}f_{A}$~
& $~~~~12 f_{ V}~~~$ & $~~~~ 12 f_{A}~~~$
\\ \hline \hline
$\nu$&0&1&4&-1&6&-4 \\ \hline
$e$  &0&1&2&1&3&-1  \\ \hline
$u$  &0&1&2&1&3&-1  \\ \hline
$d$  &0&1&-2&1&-3&-1 \\ \hline
\end{tabular}
\vskip 0.1cm
{\bf Table 1}:  U(1)$'$ charge assignment for the standard model fermions

\end{center}

After the electroweak symmetry breaking, 
the gauge sector of the Lagrangian with $Z'$ boson is given by
\be
{\cal L}_{\rm gauge}={\cal L}_{\rm kinetic}
              +{\cal L}_{\rm mass}+{\cal L}_{\rm mix} ,
\ee 
where the kinetic term and the mass term are written as
\bea
{\cal L}_{\rm kinetic} &=& - {1 \over 4 } 
          ( \hat{F}^{\mu \nu} \hat{F}_{\mu \nu}
             + Z^{\mu \nu} Z_{\mu \nu} 
             + {Z^{\prime}}^{\mu \nu} {Z^\prime}_{\mu \nu}), \\
{\cal L}_{\rm mass} &=& { 1 \over 2} 
          ({m_Z}^2 Z^{\mu} Z_{\mu} 
           + {m_{Z^\prime}}^2 {Z^\prime}^\mu {Z^\prime}_\mu ) ,  
\eea
where $\hat{F}$, $Z_{\mu \nu}$, and $Z'_{\mu \nu}$
are the usual field strength tensor for the fields
$\hat{A}_\mu$, $Z_\mu$ and $Z'_\mu$ respectively. 
The fields $\hat{A}_\mu$ and $Z_\mu$ are defined by
\bea 
Z&=& c_W W_3 - s_W B , 
\nonumber \\
\hat{A}&=& s_W W_3 + c_W B, 
\eea
where the shortened notation
$s_W = \sin \theta_W$ and $c_W = \cos \theta_W$
with the weak mixing angle $\theta_W $. 
We write ${\cal L}_{\rm mix}$ including the gauge invariant 
kinetic mixing term,
\be
{\cal L}_{\rm mix}= -{\sin{\chi}\over 2} Z^{\prime}_{\mu \nu} B^{\mu \nu}
         +{\delta M }^2 Z^{\prime}_{\mu} Z^{\mu},
\ee      
where $B^{\mu \nu} = \partial_{\mu} B_{\nu} - \partial_\nu B_\mu$ 
is the field strength tensor of U(1)$_Y$ gauge boson.

The mass eigenstates ($A, Z_1, Z_2$) are obtained 
by diagonalizing the mass terms and kinetic terms
with the transformation
\be
{\left (\begin{array}{r} A~ \\
                         Z_1 \\
                         Z_2
          \end{array} \right) }
={\left (\begin{array}{rrr}  
  1 & 0 & c_W s_{\chi}  \\
  0 & c_{\zeta} & -c_{\zeta} s_W s_{\chi}+s_{\zeta} c_{\chi}  \\
  0 & -s_{\zeta} & c_{\zeta}c_{\chi} +s_{\zeta} s_W s_{\chi} 
          \end{array} \right)}
 {\left (\begin{array}{r} \hat{A} \\
                           Z~ \\
                           Z^{\prime}
          \end{array} \right) }
\ee
where
\be
\tan {2\zeta} \equiv  
    { -2 c_{\chi} (\delta M^2 +m_Z^2 s_W s_{\chi} )  
    \over
    m_{Z^\prime}^2 - m_Z^2 s_W^2 s_{\chi}^2 
                       +2 \delta M^2 s_W s_{\chi} },
\ee
where $s_{\chi}=\sin \chi$, $c_{\chi}=\cos \chi$, 
and
$s_{\zeta}=\sin \zeta$, $c_{\zeta}=\cos \zeta$. 
The lighter $Z_1$ boson is identified to the ordinary $Z$ boson. 
We recast the Lagrangian in terms of the mass eigenstates $Z_1$ and $Z_2$ 
to obtain the interaction terms of $Z_i f \bar f$, $i=1,2$ as
\bea
-{\cal L}_{Z_i f \bar f } 
&=& {e \over  2 s_W c_W} 
       \left[ \left(1 + {\alpha T \over 2} \right)
          \sum_{f} \bar{\Psi}_f \gamma ^\mu
       [ (g^f_V + \zeta \tilde{f}^f_V )
                   -(g^f_A + \zeta \tilde{f}^f_A )\gamma_5 ]
             {\Psi}_f {Z_1}_{\mu}  \right.
\nonumber \\
&&~~~~~~~~~~~~ \left. + \sum_{f} \bar{\Psi}_f \gamma^\mu
             [ (h^f_V - \zeta g^f_V )
                   -(h^f_A - \zeta g^f_A) \gamma_5 ] 
             \Psi_f {Z_2}_{\mu} \right]
\nonumber \\
&\equiv&  \sum_{f} \sum_i {\bar{\Psi}}_f {\gamma ^\mu}
             [ {V_i}^f -{A_i}^f {\gamma}_5 ] 
             {\Psi}_f {Z_i}_{\mu},  
\eea
where the SM couplings are modified by $Z$-$Z'$ mixing effects 
\be 
g^f_A = T^f_3,~~~~ g^f_V =T^f_3 - 2 Q^f s_*^2 ,
\ee
and the extra U(1) couplings 
\bea
\tilde{f}^f_{V,A} &\equiv& {g^\prime \over g} {f^f_{V,A} \over \cos \chi} ,
\nonumber \\
{h^f}_V &=& {{\tilde{f}}^f}_V + \tilde{s} ( {T^f}_3 - 2Q^f )\tan\chi,  
\nonumber \\
{h^f}_A  &=& {{\tilde{f}}^f}_A + \tilde{s} {T^f}_3 \tan\chi .
\eea
Effective weak mixing angles are defined by
\bea 
s_{*}^2 &=& s_W^2 + \zeta~ c_W^2 s_W \tan \chi
                -\zeta^2 { c_W^2 s_W^2  \over c_W^2 - s_W^2 }
                 \left({M_2^2 \over M_1^2} -1 \right), 
\nonumber \\ 
\tilde{s} &=& s_W 
          + {s_W^3 \over  c_W^2-s_W^2 }
      \left( {\alpha S \over 4 c_W^2 } - {\alpha T \over 2} \right),
\eea
where the Peskin-Takeuchi variable $S$ and $T$ \cite{peskin} 
are given by
\bea 
  \alpha S &=& 4 \zeta {c_W}^2 s_W \tan\chi ,
\nonumber \\
  \alpha T &=& \zeta^2 \left( {M_2^2 \over M_1^2}  -1 \right) 
                +2\zeta s_W \tan\chi,
\eea   
up to the leading order of $\zeta$.

\section{Top quark pair production in the off-diagonal basis}

For the process
\be \label{process}
e^- (p_1 , s_1 )~~ e^+ (p_2 , s_2 ) \rightarrow
t   (k_1 , r_1 )~~ {\bar t}(k_2 , r_2 ),
\ee
we have s-channel Feynman diagrams mediated by
photon, $Z$ and $Z'$ boson exchanges depicted in Fig. 1.
In Eq. (\ref{process}), $p_i$ and $k_i$ denote the momenta 
and $s_i$ and $r_i$ the polarizations
of electrons and top quarks respectively. 
In the center of momentum (CM) frame, we write the momenta as
\be
p_1 = (E , E \hat{\bf n} ),~~~ p_2 = (E,- E \hat{\bf n}),~~~~
k_1 = (E, 0,0,|{\bf k}| ),~~~ k_2 = (E, 0,0,- |{\bf k} |),
\ee   
where the unit vector $\hat{\bf n}=(-\sin \theta , 0 , \cos \theta )$ 
indicates the spatial direction of the electron beam.
We assume that the whole process is confined on the $xz$-plane 
when taking the direction of the produced top quark
to be $z$-axis.

We study the spin configuration of top quark pair
in a generic spin basis suggested in Ref. \cite{parke1}.
The spin states of the top quark and top anti-quark
are defined in their own rest-frame by decomposing their spins
along reference axes.
The reference axis for top quark is
expressed by an angle $\xi$ between the axis and 
the top anti-quark momentum in the rest frame of the top quark
as depicted in Fig. 2.
The usual helicity basis is obtained by taking $\xi=\pi$.
In this general spin basis, 
the explicit expression for spin four-vectors of the $t \bar{t}$ 
is given by : 
\bea
r_1 &=& ( -{|{\bf k}| \over m} \cos \xi,\sin \xi,0, -{E \over m} \cos \xi ),
\nonumber \\
r_2 &=& ( -{|{\bf k}| \over m} \cos \xi,-\sin \xi,0,{E \over m} \cos \xi ), 
\eea
in the CM frame.
It is to be notified that the spin vectors of the produced top quark pair 
lie in the production plane at tree level 
if the CP-invariance of the scattering amplitude is preserved.

We have the scattering amplitudes for each spin configurations 
of top quark pair produced by the left-handed polarized electron 
and right-handed polarized positron beams, including $Z^\prime$ effects, 
\bea \label{LR}
{\cal M}(LR,\uparrow \uparrow)&=&
- C_1 s (\cos {\theta} \sin{\xi} - 1/\gamma \sin{\theta} \cos{\xi})
                     + C_2 s \beta \sin{\xi}  
           =-{\cal M}(LR,\downarrow \downarrow), 
\nonumber \\
{\cal M}(LR,\downarrow \uparrow)&=& 
C_1 s (\cos {\theta} \cos{\xi} +1 + 1/\gamma \sin{\theta} \sin{\xi})
                - C_2 s \beta (\cos{\theta}+\cos{\xi})  
\nonumber \\
{\cal M}(LR,\uparrow \downarrow)&=& 
C_1 s (\cos {\theta} \cos{\xi} -1 + 1/\gamma \sin{\theta} \sin{\xi})
                + C_2 s \beta (\cos{\theta}-\cos{\xi}). 
\eea
where $\beta \equiv \sqrt{1-4{m_t}^2 /s}$ 
and $\gamma \equiv 1/\sqrt{1-{\beta}^2}$.
The coefficients $C_1$ and $C_2$ are defined 
by $ C_1 = P_{VV} +P_{VA}$ and $C_2 = P_{AV} +P_{AA}$, 
and the effective coupling strength of 
current-current interactions $P_{\alpha \beta}$ are given by 
\bea
P_{VV}/ \sqrt{N_{c}}&\equiv&e^2 Q_t Q_e {D_0}(s) + {V_1}^t {V_1}^e 
                   {D_1}(s)+{V_2}^t {V_2}^e {D_2}(s) 
\nonumber \\
P_{VA}/ \sqrt{N_{c}}&\equiv&-{V_1}^t {A_1}^e {D_1}(s) 
                  - {V_2}^t {A_2}^e {D_2}(s) 
\nonumber \\
P_{AV}/ \sqrt{N_{c}}&\equiv&-{A_1}^t {V_1}^e {D_1}(s) 
                  - {A_2}^t {V_2}^e {D_2}(s)
\nonumber \\
P_{AA}/ \sqrt{N_{c}}&\equiv&{A_1}^t {A_1}^e {D_1}(s) 
                  + {A_2}^t {A_2}^e {D_2}(s),
\eea
where $N_{c}$ is the number of colors and 
$Q_{t(e)}$ is the electric charge for the top quark (electron).
$D_0$, $D_1$ and $D_2$ are the propagation factors 
for the photon, $Z_1$ and $Z_2$ bosons respectively; 
\be
D_0 (s) \equiv \frac{1}{s},~~~~
D_1 (s) \equiv \frac{1}{s-m_1 ^2},~~~~ 
D_2 (s) \equiv \frac{1}{s-m_2 ^2 },
\ee
while
$V_i ^f $ and $A_i ^f$ are the model-dependent 
vector and axial vector couplings for fermion $f$ 
and gauge boson $i=$ photon, $Z_1$, $Z_2$,
defined in Eq. (10).  

The scattering amplitudes for right-handed polarized electron 
and left-handed polarized positron are obtained in the similar manner
\bea \label{RL}
 {\cal M}(RL,\uparrow \uparrow)
           &=& C_1 s 
  (\cos {\theta} \sin{\xi} - 1/\gamma \sin{\theta} \cos{\xi})
               + C_2 s \beta \sin{\xi}
           = -{\cal M}(RL,\downarrow \downarrow), 
\nonumber \\
 {\cal M}(RL,\downarrow \uparrow)
  &=& -C_1 s 
  (\cos {\theta} \cos{\xi} -1 + 1/\gamma \sin{\theta} \sin{\xi})
                   + C_2 s \beta (\cos{\theta}-\cos{\xi}),  
\nonumber \\
 {\cal M}(RL,\uparrow \downarrow)
  &=& -C_1 s (\cos {\theta} \cos{\xi} +1 + 1/\gamma \sin{\theta} \sin{\xi})
                   - C_2 s \beta (\cos{\theta}+\cos{\xi}), 
\eea 

There exist the angles $\xi_{L}$ and $\xi_{R}$ such that 
the scattering amplitudes for the like-spin states of top quark pair,
$(\uparrow ,\uparrow)$ and $(\downarrow , \downarrow)$
vanish for the left- and right-handed electron beam, respectively.
>From the Eq. (\ref{LR}) and (\ref{RL}),
we find the angle $\xi_L$ and $\xi_R$ 
\bea
{\xi}_L(s,\theta) &\equiv& \arctan \left[ \frac{\tan \theta}
                   {\gamma (1-(C_2/ C_1) \beta \sec \theta)} \right],
\nonumber \\
{\xi}_R(s,\theta) &\equiv& \arctan \left[ \frac{\tan \theta}
                   {\gamma (1+(C_2/ C_1) \beta \sec \theta)} \right],
\eea   
which is always defined in terms of the scattering angle $\theta$.
It is called the {\it off-diagonal basis} 
since only the scattering amplitudes for off-diagonal spin states 
are non-zero \cite{parke1}.
For given kinematics, the angles ${\xi}_{L,R}$ are determined 
by the model-dependent ratio $(C_2/C_1)$,
so ${\xi}_{L,R}$ depend upon the existence of new physics.

One more interesting feature of the off-diagonal basis is that
the process into the $(\uparrow \downarrow)$ state 
for the left-handed electron beam and the 
$(\downarrow \uparrow)$ state for the right-handed one
is dominant.
This pure dominance is very stable under the one-loop QCD corrections 
where the soft gluon emissions dominate
so that the QCD corrections are factored out.
At high energy, the degree of this dominance is close to
100\%\cite{Why}.

\section{Analysis }

In the off-diagonal basis,
the scattering amplitudes of the like-spin states are 
identical to zero and so are the corresponding cross sections.
Including new physics effects,
the basis does not remain as the off-diagonal basis any more 
and the characteristic features of the off-diagonal basis
is modified through the model-dependence of the angle $\xi_{L,R}$.
The $Z'$ boson exchange diagrams yield the deviation of
the cross sections for like-spin states from zero.
Therefore observation of sizable cross sections
for like-spin states can be a smoking-gun signals of new physics.

In Fig. $3-5$, we plot the differential cross sections
for left-handed polarized electron beam
with respect to the scattering angle 
in the $\psi$-, $\chi$- and $\eta$-models respectively.
The SM predictions are denoted by solid lines.
The dashed lines denote the model predictions with
no kinetic mixing terms, the dotted lines 
the predictions with the kinetic mixing $\tan \chi = 0.2$
and the dash-dotted lines with $\tan \chi = -0.2$.
For the numerical analysis, we take $m_{Z'}=600$ GeV,
the lower bound from the direct search by CDF \cite{CDF},
and the mixing angles to be the latest bounds in Ref. \cite{erler}
to maximize the new contributions.
We have 2 lines for each predictions
corresponding to the upper and lower limits
of the $Z-Z'$ mixing angle $\zeta$ respectively.

It is apparent from the figures that the cross sections 
$\sigma(\uparrow \uparrow )$ and $ \sigma( \downarrow \downarrow)$ 
are nonzero with $Z'$ boson effects, 
which can be as large as  $10^{-2}$ pb,
of order 1\% of the total cross section of $t \bar{t}$ production.
With the expected integrated luminosity
$\int {\cal L} > 50$ fb$^{-1}$ 
for energy at $\sqrt{s} = 500$ GeV, 
we will have more than 500 events for like-spin states,
which is sufficient to examine the nonzero cross section.
We also find that the pure dominance of $(\uparrow \downarrow)$ state 
is contaminated with $Z'$ boson effects from the figures.
Cross sections for states other than $(\uparrow \downarrow)$ state
increases with $Z'$ boson effects in general.
However the pure dominance is essentially affected by 
the alteration of  $\sigma(\uparrow \downarrow)$, 
since actually the total cross section is still dominated by
$(\uparrow \downarrow)$ state.
We present the ratios of $\sigma(\uparrow \downarrow)/\sigma_{total}$
for each model in Table 2. 

Since the asymmetry of vector and axial vector charges to
$Z'$ boson features the models, the forward-backward
asymmetry of $t \bar{t}$ production can be 
an useful observable to discriminate models.
The forward-backward asymmetry $A_{FB}$ defined by
\be
A_{FB} = \frac{\sigma_F - \sigma_B} 
              {\sigma_F + \sigma_B}
\ee
increases with $Z'$ boson effects in the $\psi$-model
while it decreases in the $\chi$- and $\eta$-model 
from the figures.
The $\eta$-model is a mixture of $\psi$- and $\chi$-models
and behaves about halfway.
The smallness of $A_{FB}$ is a characteristic feature of $\chi$-model.
The asymmetry $A_{FB}$ for each model is listed in Table 2.

We find that the kinetic mixing derives much shift on the observables.
Here the kinetic mixing is taken to be $\tan \chi = \pm0.2$,
which is the bound obtained in Ref. \cite{rizzo}.
It is to be notified that the effects of kinetic mixing term
act on the effects of $Z'$ boson 
both constructively and destructively 
with respect to the sign of $\tan \chi$.
The effects of $Z'$ boson may be diluted and even canceled
by the kinetic mixing effects.
For instance, the pure dominance of $(\uparrow \downarrow)$
final state is almost recovered in the $\psi$-model
when $\tan \chi =-0.2$.
In this case, the precise measurement of $A_{FB}$
can still be an evidence of $Z'$ boson.
Hence it is essential to perform the analysis 
with more than 2 observables 
to probe the $Z'$ effects and to discriminate the models.

\begin{center}
\begin{tabular}{cccccccc}
\hline \hline
 & Models & $\tan \chi$ & 
 & $\sigma(e^-_L e^+_R \to t_{\uparrow} \bar{t}_\downarrow)/\sigma_{total}$ 
 & & $A_{FB}$ & \\
\hline \hline
 & SM & & & 99.3\% & & 0.4046 & \\
\hline
 &    & 0 & & 98.5\% & & 0.5539 & \\
 & $\psi$ & 0.2 & & 95.6\% & & 0.6796 & \\
 &    & -0.2 & & 99.2\% & & 0.4703 & \\
\hline
 &    & 0 & & 94.4\% & & 0.0382 & \\
 & $\chi$ & 0.2 & & 95.4\% & & 0.0886 & \\
 &    & -0.2 & & 93.6\% & & 0.0056 & \\
\hline
 &    & 0 & & 98.2\% & & 0.2556 & \\
 & $\eta$ & 0.2 & & 97.5\% & & 0.2036 & \\
 &    & -0.2 & & 98.9\% & & 0.3279 & \\
\hline
\hline
\end{tabular}
\end{center}
\vskip 0.3cm
{\bf Table 2}: The ratios of the cross section for 
$(\uparrow \downarrow)$ spin state of top quark pair 
to the total cross section of $e^-_L e^+_R \to t \bar{t}$ production
in the SM off-diagonal basis
and the forward-backward asymmetries are presented 
for the standard model, $\psi$-, $\chi$-, and $\eta$-models.

\section{Summary and conclusions}

We have explored the effects of $Z'$ boson arising 
in the supersymmetric E$_6$ model framework 
at $e^- e^+ \to t \bar{t}$ process,
including the kinetic mixing terms.
Considering the spin configuration of produced top quark pair,
we propose useful probes not only to search for the $Z'$ boson
but also to discriminate the models corresponding to 
the pattern of gauge group decomposition.
Provided that we take the off-diagonal spin basis of the SM,
the existence of nonzero cross sections for diagonal spin states,
$t_\uparrow \bar{t}_\uparrow$ and $t_\downarrow \bar{t}_\downarrow$
can be a direct evidence of new physics.
As a matter of fact, only one spin configuration is appreciable
for top quark pair in this basis
and violation of such a pure dominance of a peculiar spin state 
is a signature of $Z'$ gauge boson, which is almost 
free from loop corrections.
Alternatively, $A_{FB}$ is an effectual observable
to probe $Z'$ boson due to the asymmetry of
left- and right-handed couplings of $Z'$ boson to fermions.
Meanwhile it is shown that the kinetic mixing effects results in
substantial shift on the observables discussed here.
Moreover changing the sign of kinetic mixing term, its effects
can be additive or subtractive to the mass mixing effects,
by which it is possible that the $Z'$ effects is wiped out.
As a consequence, we conclude that combined analysis
with more than one observable is indispensable 
to study the structure of $Z'$ gauge boson.

\acknowledgments

This work is supported in part
by the BK21 program of Ministry of Education
for Seoul National University
and by the Korea Research Foundation (KRF-2000-D00077).

%


\newpage
{\Large \bf Figure Captions}
\vskip 2cm

\begin{description}

\item
Fig. 1 :
Feynman diagrams for the $e^- e^+ \to t \bar{t}$ process
(a) in the standard model, (b) with the $Z'$ boson.

\item
Fig. 2 :
Definition of generic top spin basis 
(a) in the top quark rest frame, 
(b) in the anti-top quark rest frame.

\item
Fig. 3 :
The differential cross sections in the $\psi$-model 
for the spin configuration of the top quark pair, 
drawn with respect to the scattering angle of the top quark 
at $\sqrt{s} = 500$ GeV with the left-handed electron beam.
The solid line denotes the SM prediction; the dashed line includes
the $Z'$ boson effects with the kinetic mixing $\tan \chi = 0$,
the dotted line with $\tan \chi = 0.2$,
the dash-dotted line with $\tan \chi = -0.2$.

\item
Fig. 4 :
The differential cross sections in the $\chi$-model 
for the spin configuration of the top quark pair, 
drawn with respect to the scattering angle of the top quark 
at $\sqrt{s} = 500$ GeV with the left-handed electron beam.
The solid line denotes the SM prediction; the dashed line includes
the $Z'$ boson effects with the kinetic mixing $\tan \chi = 0$,
the dotted line with $\tan \chi = 0.2$,
the dash-dotted line with $\tan \chi = -0.2$.

\item
Fig. 5 :
The differential cross sections in the $\eta$-model 
for the spin configuration of the top quark pair, 
drawn with respect to the scattering angle of the top quark 
at $\sqrt{s} = 500$ GeV with the left-handed electron beam.
The solid line denotes the SM prediction; the dashed line includes
the $Z'$ boson effects with the kinetic mixing $\tan \chi = 0$,
the dotted line with $\tan \chi = 0.2$,
the dash-dotted line with $\tan \chi = -0.2$.

\end{description}

\begin{center}
\begin{figure}[htb]
\hbox to\textwidth{\hss\epsfig{file=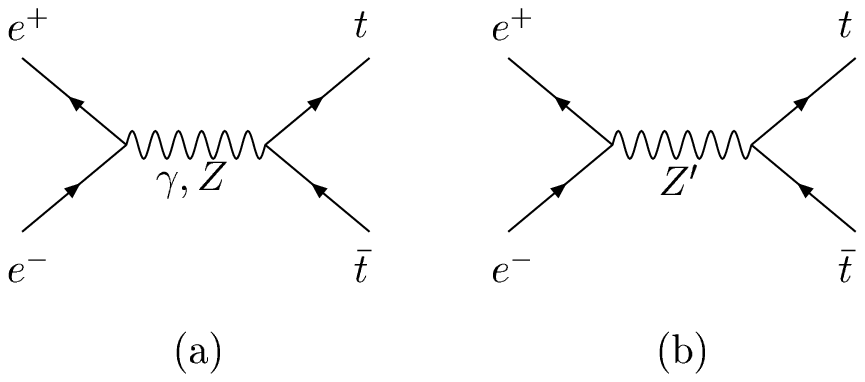,width=15cm}\hss}
\vspace{1cm}
\caption{\it 
}
\end{figure}
\end{center}

\newpage
\begin{center}
\begin{figure}[htb]
\hbox to\textwidth{\hss\epsfig{file=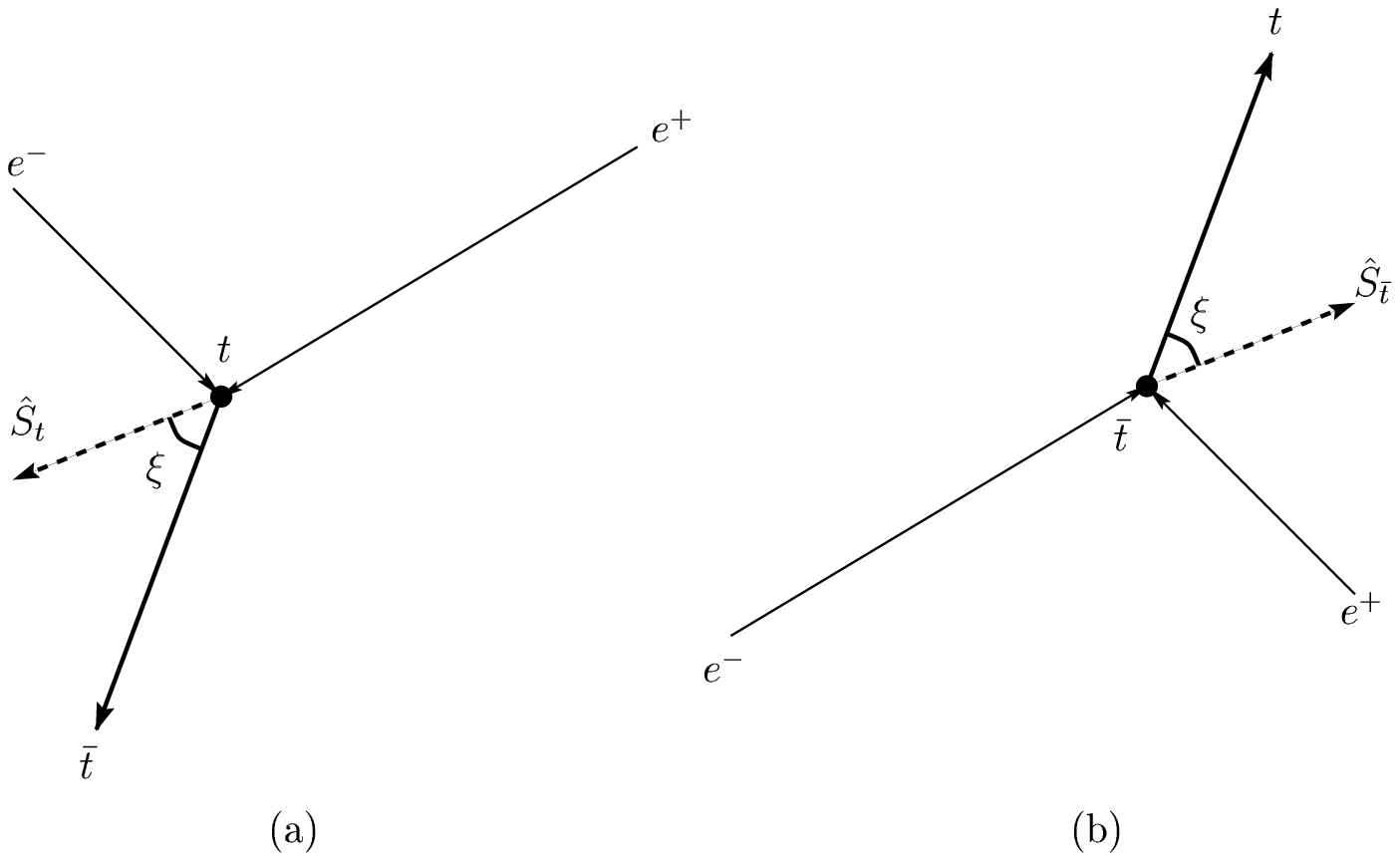,width=15cm}\hss}
\vspace{1cm}
\caption{\it 
}
\end{figure}
\end{center}

\newpage
\begin{center}
\begin{figure}[htb]
\hbox to\textwidth{\hss\epsfig{file=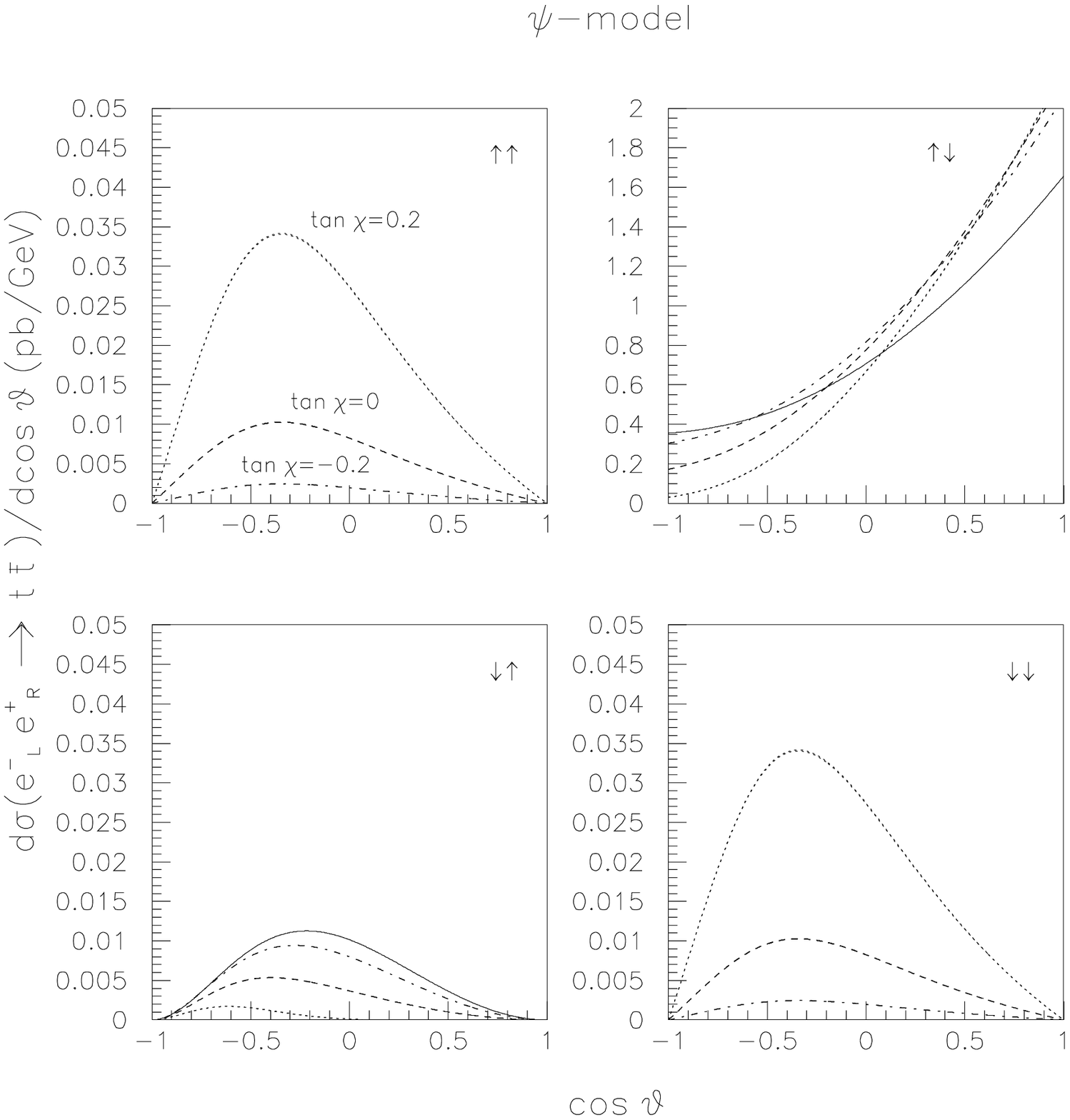,height=15cm}\hss}
\vspace{1cm}
\caption{\it 
}
\end{figure}
\end{center}

\newpage
\begin{center}
\begin{figure}[htb]
\hbox to\textwidth{\hss\epsfig{file=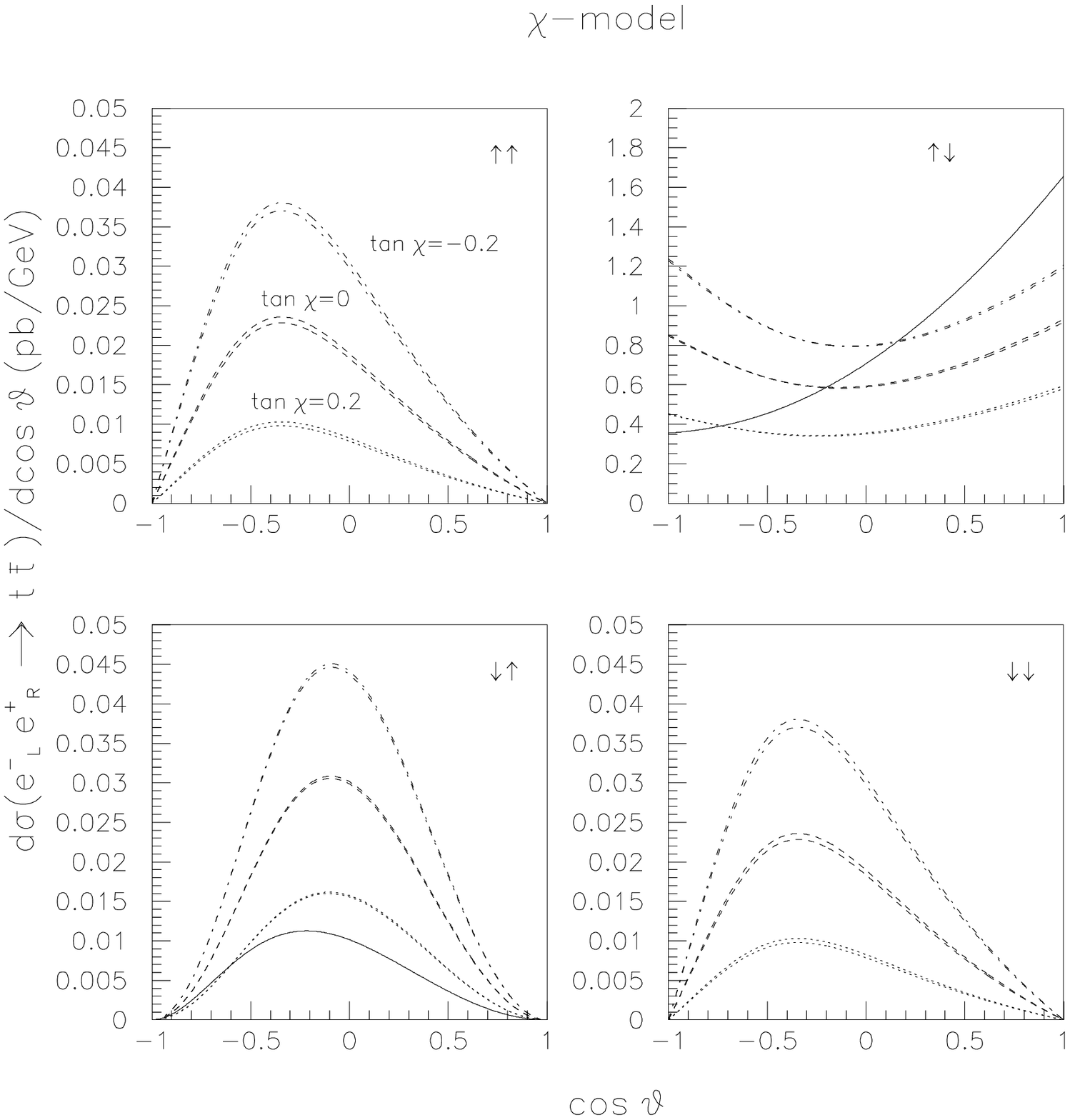,height=15cm}\hss}
\vspace{1cm}
\caption{\it 
}
\end{figure}
\end{center}

\newpage
\begin{center}
\begin{figure}[htb]
\hbox to\textwidth{\hss\epsfig{file=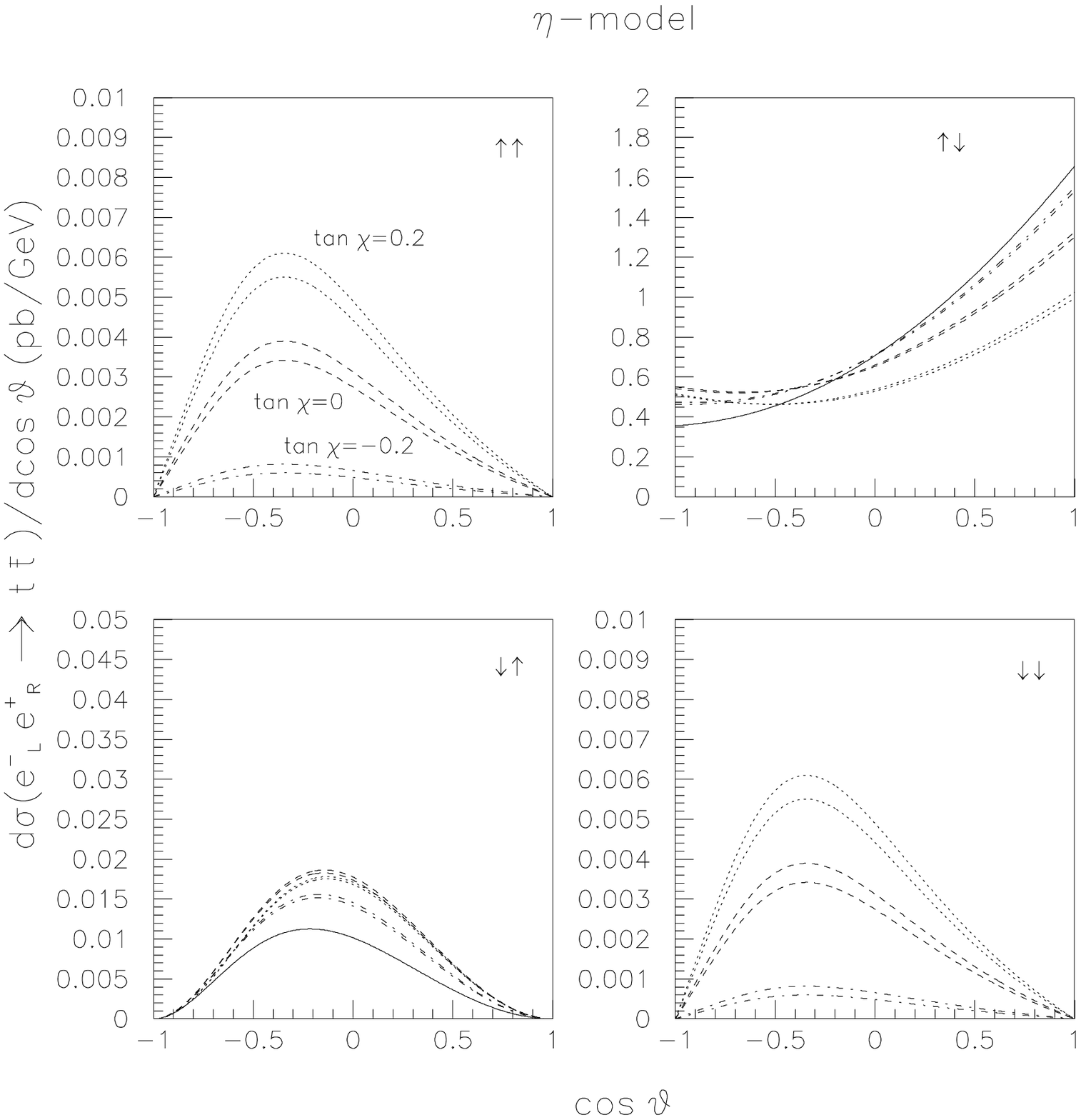,height=15cm}\hss}
\vspace{1cm}
\caption{\it 
}
\end{figure}
\end{center}

\end{document}